\documentclass[onecolumn,showpacs,multicol,amsmath,amssymb]{revtex4-2}
\usepackage{amssymb}
\usepackage{amsmath}
\usepackage{graphicx}
\usepackage{epstopdf}
\usepackage{bm}
\usepackage{gensymb}

\newcommand{\be}{\begin{equation}}
	\newcommand{\ee}{\end{equation}}
\newcommand{\kk}{\bf{k}}

\newcommand{\bea}{\begin{eqnarray}}
	\newcommand{\eea}{\end{eqnarray}}

\newcommand{\bd}{\begin{displaymath}}
	\newcommand{\ed}{\end{displaymath}}
\newcommand{\ba}{\begin{array}}
	\newcommand{\ea}{\end{array}}
\newcommand{\bi}{\begin{itemize}}
	\newcommand{\ei}{\end{itemize}}
\newcommand{\bc}{\begin{center}}
	\newcommand{\ec}{\end{center}}
\newcommand{\bfl}{\begin{flushleft}}
	\newcommand{\efl}{\end{flushleft}}
\newcommand{\bfr}{\begin{flushright}}
	\newcommand{\efr}{\end{flushright}}

  \def\bd{{\bf d}}

\def\6{\partial}

\def\={\!\!\!&=&\!\!\!}
\def\+{\!\!\!&&\!\!\!+~}
\def\-{\!\!\!&&\!\!\!-~}

\newcommand\redout{\bgroup\markoverwith{\textcolor{red}{\rule[.5ex]{2pt}{0.4pt}}}\ULon}
\usepackage[colorlinks=true,citecolor=blue]{hyperref}
\hypersetup{colorlinks=true,citecolor=blue,linkcolor=red,urlcolor=blue}

\begin{document}

\title{
	Unveiling Non-Kitaev Interactions and Field-Angle Dependence in Topological Magnon Transport of $\alpha$-RuCl$_3$
}

\author{Hamid mosadeq}\email{mosadegh@sku.ac.ir}
\affiliation{Department of physics, Faculty of Science, Shahrekord university, Shahrekord 88186-34141, Iran}

\author{Mohammad-Hossein Zare}\email{zare@qut.ac.ir}
\affiliation{Department of Physics, Qom University of Technology, Qom 37181-46645, Iran}


\begin{abstract}
	
	Honeycomb lattice Kitaev magnets exhibit exotic magnetic properties governed by the Kitaev interaction.
	This study delves into $\alpha$-RuCl$_3$, a prototypical example described by effective Hamiltonians encompassing bond-dependent Kitaev interactions alongside additional terms such as the Heisenberg interaction and symmetric off-diagonal exchange interactions.
	These non-Kitaev terms significantly influence $\alpha$-RuCl$_3$'s low-temperature magnetism, impacting both magnetic order and excitations.
	We employ spin-wave theory to elucidate the topological nature of magnetic excitations within the polarized state of $\alpha$-RuCl$_3$ under an external magnetic field.
	Our focus lies on transverse magnon conductivities, specially the thermal Hall conductivity and spin Nernst coefficient.
	The calculations unveil a pronounced dependence of the magnitude and sign structure of the low-temperature transverse thermal conductivities on both the applied magnetic field's orientation and the exchange parameters within the nearest neighbor Heisenberg-Kitaev-Gamma-Gamma$'$ $(JK\Gamma\Gamma')$ model, which govern the nature and strength of spin interactions.
	This theoretical framework facilitates critical comparisons with experimental observations, ultimately aiding the identification of an effective Hamiltonian for Kitaev magnets exemplified by $\alpha$-RuCl$_3$.
	
\end{abstract}
	
	\flushbottom
	\maketitle

	\section*{Introduction}
	The exactly solvable spin-1/2 Kitaev honeycomb model of bond-dependent Ising-type interactions is an alternative pathway to a quantum spin liquid (QSL) as its ground state, in which frustrations originated from the bond-directional interactions~\cite{Kitaev:2006}.
	Frustrations play a key role in inducing quantum fluctuation and suppressing of long-range magnetic order down to absolute zero temperature. 
	The Kitaev QSL state possess two fractional quasiparticles from the fractionalized of spins: itinerant Majorana fermionic quasiparticles and localized $\mathbb{Z}_2$ fluxes~\cite{Savary:2017,Do:2017,knolle:2019}. 
	Experimental identification of exotic fractionalized excitations is half quantized thermal Hall conductivity at low-energies~\cite{Kane:1997}. 
	Majorana fermions are indeed promising candidates for advancing the field of topological quantum computation due to it highly resistant against certain kinds of errors that plague conventional quantum computers~\cite{Kitaev:2006,kitaev2001unpaired,kitaev2003fault}.
	
	The magnetic properties of $\alpha$-RuCl$_3$, a promising candidate for the realization of Kitaev quantum spin liquids, have been extensively investigated~\cite{Kitaev:2006,Jackeli:2009}.

\begin{figure}[t]
		\centerline{\includegraphics[width=0.95\columnwidth]{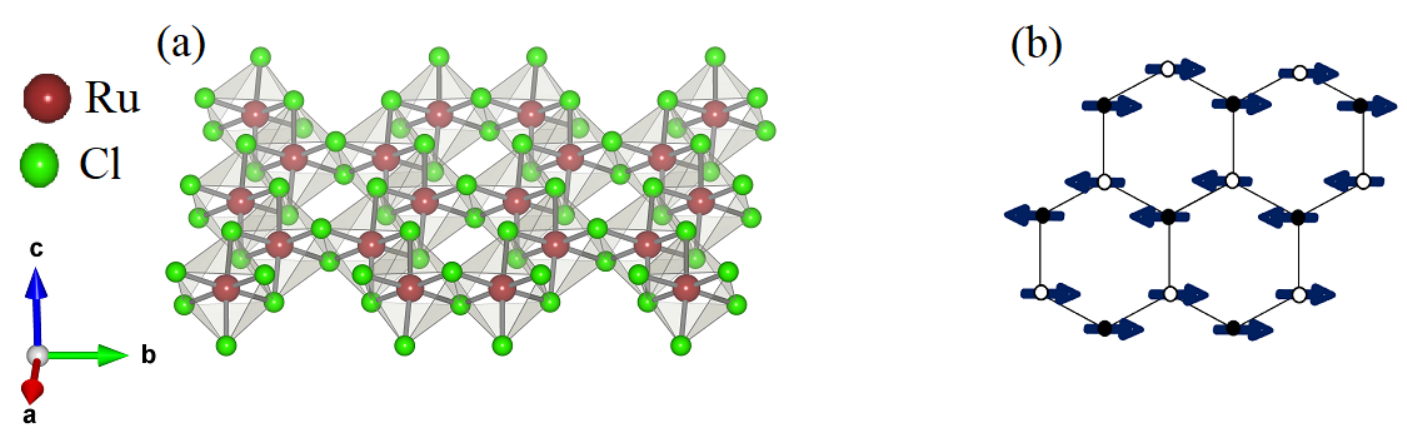}}
		\caption{(a) Crystal structure of $\alpha$-RuCl$_3$ compound with
			the Ru$^{3+}$ cations (with an electronic configuration: 4d$^{5}$) forming a  honeycomb lattice 
			and surrounded by six Cl$^{-}$ anions, forming edge-sharing octahedra.
			(b) Schematic representation of spin configuration of Ru$^{3+}$ atoms in a zigzag antiferromagnetic phase of  $\alpha$-RuCl$_3$. }
		\label{fig:fig1}
	\end{figure}

	$\alpha$-RuCl$_3$ is a layered antiferromagnet (AFM), with Ru atoms arranged in a honeycomb lattice and surrounded by six chlorine ions in octahedra symmetry (Fig.~\ref{fig:fig1}a).
	In zero magnetic field, the magnetic structure in the honeycomb layers of ruthenium chloride $\alpha$-RuCl$_3$ is a zigzag AFM order below $T_N \sim 7-8$ K~\cite{Sears:2015,Cao:2016,Sears:2017} (Fig.~\ref{fig:fig1}b).
	It is worth noting that non-Kitaev terms such as the Heisenberg $J$ and symmetric off-diagonal $\Gamma^{(')}$ exchange interactions play a predominant role for stabilizing of the low-temperature zigzag AFM order, driving the compound away from the QSL phase~\cite{Sears:2015,Johnson:2015,Banerjee:2017,kubota2015successive}. 
	In the context of first-principle analysis, the additional non-Kitaev terms can originate from various sources. For instance, $\Gamma$ and $\Gamma'$ can emerge due to spin-orbit coupling~\cite{Rau:2014} and trigonal distortion~\cite{Winter:2016,kim:2016crystal},
	respectively.
	Theoretical investigations suggest that the spin interactions within $\alpha$-RuCl$_3$ can be experimentally manipulated through modifications in layer stacking and octahedral distortion~\cite{luo2022interplay,arakawa2021floquet,strobel2022comparing,sriram2022light,kumar2022floquet}.
	Experimental observations point out that the zigzag order can be suppressed by a moderate in-plane magnetic field (about 7 T) and the possibility to induce an intermediate QSL state before the polarized state at higher fields~\cite{Balz:2019,Wellm:2018,Kasahara:2018,yokoi:2021half,Czajka:2021,ahmadi2024ground}.

	Discovered two-dimensional (2D) ferromagnetic and antiferromagnetic materials, which are often van der Waals materials,
	have attracted exotic attention due to their unique properties and potential applications in spintronics and quantum technologies.
	This class of materials provide a great playground for realizing intriguing topological phenomena, including the emergence of topologically protected magnons and the presence of magnon edge states~\cite{burch:2018,rodin:2020,rahman:2021,Li:2022a,Gibertini:2019,zare2021spin,Cong:2022,mosadeq2024frustration}. 
	For magnons, a temperature gradient could serve as an external force driving anomalous transverse magnon transport. This phenomenon manifests as the observation of thermal Hall and Nernst conductivities.
	Indeed, these transverse conductivities induced by longitudinal thermal transport can provide valuable insights to the spin excitations present in insulating quantum materials~\cite{Kane:1997,czajka:2023}.

	Recent studies have indeed delved into various lattice structures such as kagome lattice~\cite{sheikhi2021thermal,Li:2023} and honeycomb lattice~\cite{Joshi:2018,McClarty:2018,Chern:2021,Zhang:2021} to explore the realization of the topological magnon excitations.
	These studies typically involve calculating several key properties to characterize the topological nature of magnon excitations, including Berry curvature, Chern number, thermal Hall conductivity, and spin Nernst coefficient.
	Noted that the realization of magnetic insulators with high tunable Berry curvatures, Chern numbers, and transverse thermal conductivities indeed open up exciting prospects for various applications in emerging quantum technology~\cite{bao:2023}.

	Theoretical and experimental studies in Kitaev candidate magnet $\alpha$-RuCl$_3$ have revealed that the system transitions into a polarized phase at high fields. In this phase, all spins within the compound become ferromagnetically aligned in the direction of the external magnetic field ~\cite{Li:2021b,Zhou:2023,ahmadi2024ground}.
	In $\alpha$-RuCl$_3$, the critical field strength required to stabilize the polarized phase increases from the in-plane filed to the out-of-plane field due to the combining effect of strong magnetic anisotropy~\cite{kubota2015successive,chaloupka2016magnetic,yadav2016kitaev,winter2018probing} and a positive $\Gamma$ interaction~\cite{sears:2020ferromagnetic}.
	In the honeycomb magnet $\alpha$-RuCl$_3$, which exhibits a polarized state as its ground state at high fields, the thermal Hall conductivity has been observed to be strongly dependent on the direction of the magnetic field direction at low temperatures~\cite{Czajka:2021,czajka:2023,yokoi:2021half,Chern:2021,bruin:2022robustness}.
	Experimental observations and theoretical works have indeed predicted that the Kitaev interaction plays a critical role in opening up a gap in the spectrum, leading to the emergence of the topological magnon bands~\cite{Katsura:2010,Onose:2010,Zhang:2013,Hirschberger:2015,Chisnell:2015,Owerre:2016,Li:2017}. 

	Inspired by these findings, in this paper, we are interested in a theoretical investigation regarding the influence of tuning orientation an applied magnetic field and exchange parameters on transverse thermal conductivities, specially the thermal Hall conductivity and spin Nernst coefficient, in the Kitaev magnet $\alpha$-RuCl$_3$.
	We conducted a comprehensive study of topological magnon excitations in $\alpha$-RuCl$_3$.
	To accurately capture the various magnetic interactions between neighboring spins, we employed the nearest-neighbor Heisenberg-Kitaev-Gamma-Gamma$'$ $(JK\Gamma\Gamma')$ model. Furthermore, we incorporated the influence of an external magnetic field on the system.
	The purpose is focused on examining four proposed parametrizations of $\alpha$-RuCl$_3$ (see Table~\ref{table:table1}) to determine if they support topological magnons and explore how the existence of these magnons relates to the direction of the applied field within the quantum model.

	\begin{table}[b]
		\centering
		\begin{tabular}{l| cccc |l}
			\toprule
			Model & $J$ & $K$ & $\Gamma$ & $\Gamma'$  & Ref. \\
			1($JK\Gamma\Gamma'$)
			&    $-1$   &  $-8$   &  $4$   &  $-1$ & \cite{kim:2016crystal}     \\
			2($JK\Gamma\Gamma'$)
			& $-1.5$   &  $-40$   &  $5.3$   &  $-0.9$ &  \cite{suzuki:2018effective}    \\
			3($JK\Gamma\Gamma'$)
			& $0$   &  $-6.8$   &  $9.5$   &  $0$ &   \cite{ran:2017spin} \\
			4($JK\Gamma\Gamma'$)
			& $-2.5$     &  $-25$   &  $7.5$   &  $-0.5$ & \cite{Li:2021b} \\
		\end{tabular}
		\caption{
			Different exchange parameters (in meV)	of the nearest neighbor $JK\Gamma\Gamma'$ model as the minimal model for $\alpha$-RuCl$_3$. 
		}
		\label{table:table1}
	\end{table}

	\section*{Model and Method}
\begin{figure}[t]
		\centerline{\includegraphics[width=0.65\columnwidth]{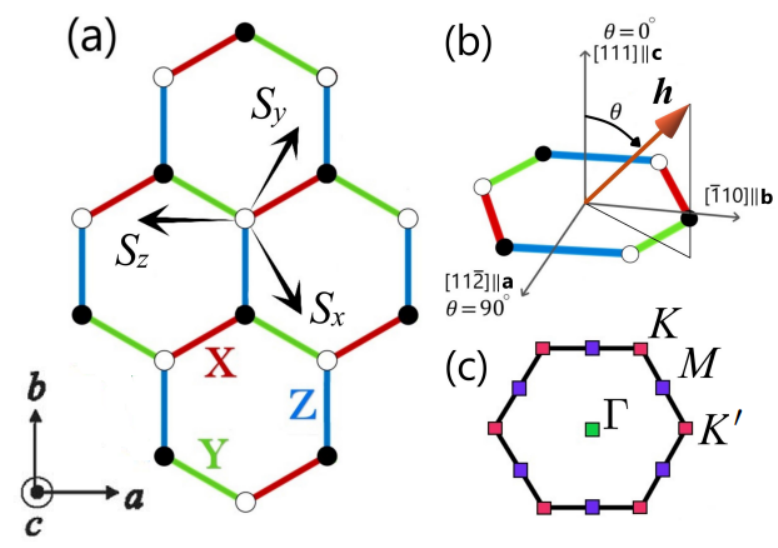}}
		\caption{(a) Schematic depiction of the Kitaev honeycomb model on a honeycomb lattice, featuring Ising-like directional interactions between Ru atoms. The three different types of Ru-Ru bonds ($x,~y,~z$) are marked by three various colors. For clarity, both the crystallographic ($abc$) and cubic ($xyz$) coordinate systems are included.
			(b) The out-of-plane [111] direction and the in-plane $[11\bar{2}]$ direction are shown with the $c$ axes and the $a$ axes, respectively. (c)
			First Brillouin zone (FBZ) of the honeycomb lattice. High-symmetry points are labeled ($\Gamma$, M, K, K$^{\prime}$).}
	\label{fig:fig2}
\end{figure}

	The $JK\Gamma\Gamma'$ model is widely recognized as the minimal model for capturing the magnetic properties of $\alpha$-RuCl3~\cite{kim:2016crystal,suzuki:2018effective,ran:2017spin,Li:2021b}.
	Within in this model, the nearest neighbor interaction between spins on a $z$-bond of the honeycomb lattice (Fig.~\ref{fig:fig2}a) reads as:
	\begin{equation}
		{H}_{ij}^{z}= J {\bf S}_{i}\cdot{\bf S}_{j}+KS_{i}^{z}S_{j}^{z}+\Gamma(S_{i}^{x}S_{j}^{y}+S_{i}^{y}S_{j}^{x})
		+\Gamma'(S_{i}^{x}S_{j}^{z}+S_{i}^{z}S_{j}^{x}+S_{i}^{y}S_{j}^{z}+S_{i}^{z}S_{j}^{y}),
		\label{eq:ham}
	\end{equation} 
	where $J$ is conventional Heisenberg interaction. $K$ represents the bond-dependent Kitaev exchange interaction. $\Gamma$ and $\Gamma'$ denote two types of symmetric off-diagonal interactions arsing from spin-orbit coupling~\cite{Rau:2014} and trigonal distortion~\cite{Winter:2016,kim:2016crystal}, respectively.
	For interactions on $\gamma=x,y$ bonds, the Hamiltonian (\ref{eq:ham}) is obtained via cyclic permutation of the spin components ($S^{x}_{i},S^{y}_{i},S^{z}_{i}$). The $JK\Gamma\Gamma'$ model under an applied external magnetic field $\bf h$ is given by:
	\begin{equation}
		{H}=\sum_{\gamma\in\{x,y,z\}}\sum_{\langle ij\rangle\in\gamma}
		H_{ij}^{\gamma}
		-{{\bf h}^{\rm T}} \sum_{i}{\bf S}_i.
		\label{eq:shh}
	\end{equation}

To elucidate the transverse thermal conductivities in $\alpha$-RuCl$_3$, we employ a well-established spin-wave theory (SWT) approach.
Building upon the theoretical framework outlined in seminal works by Refs.~\cite{holstein1940field,jones1987spin},
we analytically treat the spin Hamiltonian (\ref{eq:shh}) within the SWT formalism, thereby deriving an effective magnon Hamiltonian.
SWT is a perturbative approach used to study a collective precession of the magnetic moments around their equilibrium position refereed to as spin wave or magnons~\cite{kusminskiy2019quantum}.
In the context of describing a magnetic ground state with polarized order where all the magnetic moments (spins) are ferromagnetically aligned and the spin-quantization axis is aligned along the field direction, switching from the cubic $xyz$ coordinates to the crystallographic $abc$ coordinates involves transforming the orientation of the coordinate system to align with the crystal lattice axes, as shown in Fig.~\ref{fig:fig2}(b).
Within the reference frame defined by the spin axis, the crystallographic  $a$, $b$, and $c$ axes are represented by the basis vectors $[11{\bar 2}]$, $[{\bar 1}11]$, and $[111]$, respectively~\cite{rusnavcko2019kitaev,chaloupka2015hidden}.
In the cubic coordinate, we employ two angles, $\theta$ and $\phi$, to specify the spin orientation in the polarized phase as
$(S_x,S_y,S_z)=S(\sin\theta\cos\phi,\sin\theta\sin\phi,\cos\theta)$.
To represent the spin rotation, we introduce a rotation matrix $R$. This matrix will be defined as follows~\cite{jones1987spin}:

\begin{equation} \label{rotationmatrix}
	R_{}=\left[\begin{matrix}\cos\theta_{i}\cos\phi_{i} & -\sin\phi_{i} & \sin\theta_{i}\cos\phi_{i}\\
		\cos\theta_{i}\sin\phi_{i} & \cos\phi_{i} & \sin\theta_{i}\sin\phi_{i}\\
		-\sin\theta_{i} & 0 & \cos\theta_{i}
	\end{matrix}\right]
\end{equation}

For a magnetic field $({\bf h})$ aligned along the [001] axis, the corresponding angles $(\theta,\phi)$ in this rotation matrix are equivalent to $(0,\phi)$. When ${\bf h}$ is parallel to the $[111]$ axis, $(\theta,\phi)$ becomes $(\cos^{-1}(\sqrt{1/3}),\pi/4)$. Similarly, when ${\bf h}$ is aligned with the $[11\bar{2}]$ axis, $(\theta,\phi)$ is equal to $(\pi-\sin^{-1}(\sqrt{2/3}),\pi/4)$. Finally, if ${\bf h}$ is directed along the $[\bar{1}10]$ axis,  $(\theta,\phi)$ is given by $(\pi/2,3\pi/4)$.

By substituting ${\bf S}_{i}=R {\tilde{\bf S}}_{i}$ into the spin Hamiltonian ($\ref{eq:shh}$), we can obtain its re-expressed form in the basis of the new spin-${\tilde{S}}$ coordinate system.
Therefore, the rotated spin Hamiltonian of the $(JK\Gamma\Gamma')$ model is given by:
\begin{equation}
	H=\sum_{\gamma\in\{x,y,z\}}\sum_{\langle ij\rangle\in\gamma} 
	{\tilde{H}}^{\gamma}_{ij}
	-\sum_{i} {\tilde{\bf h}}^{\rm T}_{i}{\tilde{{\bf S}}}^{}_{i},
	\label{eq:rham}
\end{equation}
where ${\tilde{\bf h}}^{}_{i}=R^{\rm T}_{i}{\bf h}_{}$.  
Within in the new spin-$\tilde{S}$ basis, we employ the linearized Holstein-Primakoff transformations on the new spin operators (${\tilde{\bf S}}_i$). This involves representing these operators in terms of auxiliary bosons: $\tilde{S}_z=S-b^{\dagger}b$, $\tilde{S}^{\dagger}=\sqrt{2S}b$, $\tilde{S}^{-}=\sqrt{2S}b^{\dagger}$; where $b$ ( $b^{\dagger}$) stands for the annihilation (creation) operator of boson.
By employing Fourier transformations, we can transform the linear spin-wave theory applied to a field-polarized $(JK\Gamma\Gamma')$ model into momentum space as follows:
\begin{equation}
	H_{\text{SWT}}=\frac{S}{2}\sum_{\bf k} \Upsilon^{\dagger}_{\bf k} {\bf D}({\bf k})\Upsilon_{\bf k}	
\end{equation}
where $\Upsilon_{\bf k}=\Big(b_{1,{\bf k}},b_{2,{\bf k}},b^{\dagger}_{1,{\bf -k}},b^{\dagger}_{2,{\bf -k}}\Big)$ with $b_1$ and $b_2$ bosons living on the two sublattices of the honeycomb lattice.
${\bf D}({\bf k})$ is a  four-by-four matrix defined as:

\begin{equation} \label{eq:matrixDk}
	{\bf D}({\bf k})=\left[\begin{matrix}	{{\bf A}}({\bf k}) & { {\bf B}}({\bf k}) \\
		{ {\bf B}}^{*}({-\bf k}) & { {\bf A}}^{T}({-\bf k})
	\end{matrix}\right]
\end{equation}
where  ${{\bf A}}({\bf k})$ and ${{\bf B}}({\bf k})$ are momentum-dependent functions of the coupling constants in (\ref{eq:ham}). For more details, see Appendix.

To ensure the the commutation relation of boson is preserved, it is usually to diagonalize the matrix ${\bf D}({\bf k})$ using Bogoliubov transformations. 
Diagonalization the quadratic Hamiltonian reveals four spin-wave dispersion branches denoted by $\omega^{}_{n\bf k}$ ($n=1,...,4$).
We consider only the positive branches, discarding the negative ones, as our primary interest lies in characterizing the quasiparticle excitations within the system,  define as follows:

\begin{align}
	\omega_{1{\bf k}} &=\sqrt{\alpha_{\kk}+\beta_{\kk}}, \nonumber & \\
	\omega_{2{\bf k}} &=\sqrt{\alpha_{\kk}-\beta_{\kk}},
\end{align}
where
\begin{align}
	\alpha_{\kk} & = \lambda_0^2 + |\lambda_{1,\kk}|^2 - |\lambda_{2,\kk}|^2, \nonumber & \\ 
	\beta_{\kk} & = \sqrt{4\lambda_0^2|\lambda_{1,\kk}|^2 + (\lambda^*_{1,\kk}\lambda_{2,\kk}-\lambda_{1,\kk}\lambda^*_{2,\kk})^2}. 
\end{align}
Furthermore, the ground state energy is given by:
\begin{equation}
	E_{\rm gs}=\frac{NS^2}{2} E_{\rm cl}+E_{\rm Q},
\end{equation}
in which ${N}/{2}$ denotes the number of primitive cells, and $E_{\rm cl}$ represents the classical energy associated with magnetic fields oriented in various directions, expressed as follows:

\begin{align}
	\begin{aligned}
		{E_{\rm cl}} & =
		\begin{cases}
			3J+K-h, & \quad \text{for } {\bf h} \parallel [001], \\
			3J+K-h, &  \quad \text{for } {\bf h} \parallel [111], \\
			3J+K-\Gamma-2\Gamma'-h, &
			\quad \text{for } {\bf h} \parallel [11{\bar 2}],
			\\
			3J+K-\Gamma-2\Gamma'-h, & \quad \text{for } {\bf h} \parallel [\bar{1}10],  
		\end{cases}
	\end{aligned}
\end{align} 
and $E_{\rm Q}$ is the quantum correction to the classical energy $E_{\rm cl}$ reads as:
\begin{equation}
	E_{\rm Q}=\frac{S}{2}\sum_{k>0} [	\omega_{1{\bf k}} + 	\omega_{2{\bf k}} - 2\lambda_{0}].
\end{equation}

The application of a longitudinal temperature gradient leads to the emergence of two distinct transverse thermal currents: the heat current and the spin current.
These phenomena are recognized as  the thermal Hall effect and the spin Nernst effect, respectively.
The linear spin-wave dispersion $\omega^{}_{n\bf k}$ allows us to calculate the thermal Hall conductivity ($\kappa_{\text{TH}}^{xy}$)~\cite{matsumoto2011theoretical,matsumoto2014thermal,murakami2017thermal} and the spin Nernst coefficient ($\kappa_{\text{N}}^{xy}$)~\cite{zyuzin2016magnon,cheng2016spin} arising from magnons, as expressed in the following equations:
\begin{equation}
	\kappa^{xy}_{\text{TH}}= -\frac{k_{\rm B}^{2}T}{(2\pi)^2\hbar}\sum_{n}\int_{\rm FBZ}^{} \Big\{c_{2}[n_{\rm B}(\omega^{}_{n\bf k})]-\frac{\pi^2}{3}\Big\}\Omega_{n\bf k}^{} d{\bf k},
	\label{eq:thermal}
\end{equation}

\begin{equation}
	\kappa^{xy}_{\text{N}}=-\frac{k_{\rm B}}{(2\pi)^2}\sum_{n}\int_{\rm FBZ}^{}c_{1}[n_{\rm B}(\omega^{}_{n\bf k})] \Omega_{n\bf k}^{} d{\bf k}
	\label{eq:nernst}
\end{equation}
where FBZ signifies the first Brillouin zone, $n_{\rm B}(\omega_{n{\bf k}})=(e^{{\omega_{n{\bf k}}}/{k_{\rm B}T}}-1)^{-1}$ refers to the Bose-Einstein distribution function.
Within the context of the $k$-space band structure,  
$\Omega_{n{\bf k}}=i\epsilon_{\mu\eta}\Big[\sigma_3\frac{\partial{T_{k}^{\dagger}}}{\partial{k_\mu}}\sigma_3\frac{\partial{T_{k}^{}}}{\partial{k_\eta}}\Big]$ denotes the Berry curvature associated with the $n$th bosonic band at a specific momentum ${\bf k}$. $T_{k}$ specifically refers to a Bloch eigenstate belonging to this $n$th band at momentum ${\bf k}$.
In this context, the non-negative weights  $c_1(\tau)$ and $c_2(\tau)$ are defined as follows:
\begin{align}
	c_2(\tau) &=(1+\tau)\{\ln [(1+\tau)/\tau]\}^2 -(\ln \tau)^2 -2{\rm Li}_2(-\tau)  \nonumber \\ 
	c_1(\tau) & = (1+\tau) \ln (1+\tau) -\tau\ln\tau,
\end{align}
here, ${\rm Li}_2$ represents the dilogarithm function.
In subsequent analyses, we can disregard the constant term $-\pi^2/3$ as the sum of the Chern numbers of all bands is zero~\cite{shindou2013topological}.

\begin{figure}[t]
	\centerline{\includegraphics[width=15cm,
		height=12cm]{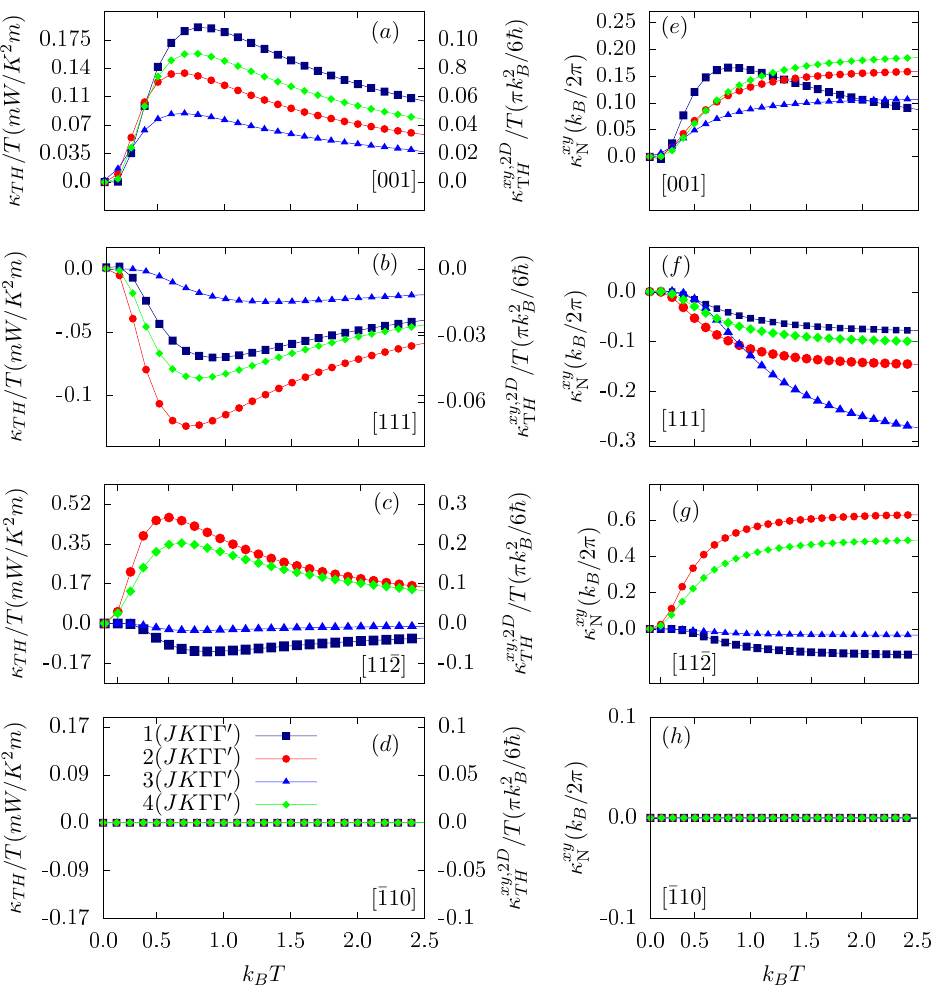}}
	\caption{
		Panels (a-d) denote the temperature dependence of the Hall conductivity ($\kappa_{\text{TH}}^{xy}$) for the models in Table~\ref{table:table1} under a variety of magnetic field directions. Field strengths $h/S|K|$ of 1.0, 0.8, 1.5, and 0.9 were applied along the [001] direction (first row), while 1.0, 1.0, 5.5, and 1.3 were applied along the [111] direction (second row). For the $[11{\bar{2}}]$ (third row) and $[{\bar{1}}10]$ (fourth row) directions, the applied field strengths were 0.18, 0.1, 1.0, and 0.1; and 0.5, 1.1, 0.1, and 0.01, respectively.
		In the right axis, the two dimensoinal thermal Hall conductivity $\kappa_{\text{\text{TH}}}^{xy,2D}/T\equiv\kappa_{\text{\text{TH}}}^{xy}d/T$ (in unit of $\pi k^{2}_{\text{B}}/6\hbar$) is included ($d=5.72 \, \textrm{\AA}$ for $\alpha$-RuCl$_3$).
		Panels (e-h) show the temperature dependence of $\kappa^{xy}_{\text{N}}$ (in unit $k_{\rm B}/2\pi$) for the same models and field directions. 
		Specific magnetic field strengths $h/{S|K|}$ were meticulously chosen to ensure the stability of the classical model in the polarized phase.}
	\label{fig:fig3}
\end{figure}

\section*{Results and Discussion}
This work delves into the ongoing debate regarding magnon thermal transport within the candidate Kitaev material $\alpha$-RuCl$_3$.
We employ a theoretical framework to analyze the temperature-dependent response of magnons, specially focusing on the alternations in the transverse thermal conductivities $\kappa^{xy}_{\text{TH}}$ and $\kappa^{xy}_{\text{N}}$ induced by a non-zero temperature gradient.
The experimental determination of these transport coefficients provides a powerful probe for elucidating the material's underlying topological character. This connection is established by linking the Chern number, a well-defined topological invariant, to the transverse thermal conductivities.
In this context, the Chern number ($\nu_n$) associated with the $n$th magnonic band is computed through a surface integral over the entire FBZ of the Berry curvature. This integral expression is given by  
$\nu_{n}=\frac{1}{2\pi}\int_{\rm FBZ} d{\bf k} \; \Omega_{n{\bf k}}$.
Notably, the sign of the Chern number ($\nu_{n}$) dictates the sign of the material's transverse thermal conductivities. 
Equations (6) and (7) define the expressions for the thermal Hall conductivity ($\kappa^{xy}_{\text{TH}}$) and the spin Nernst coefficient ($\kappa^{xy}_{\text{N}}$) via scaling factors, $c_2(\tau)$ and $c_1(\tau)$ respectively, acting on $\Omega_{n{\bf k}}$ within the integrands. 
Noted that the scaling factors $c_2(\tau)$ and $c_1(\tau)$
are non-negative weights ($c_2(\tau)\geq0$, $c_1(\tau)\geq0$), and also a minus sign appears in front of these equations.
\begin{figure}[t]
	\centerline{\includegraphics[width=1.0\columnwidth]{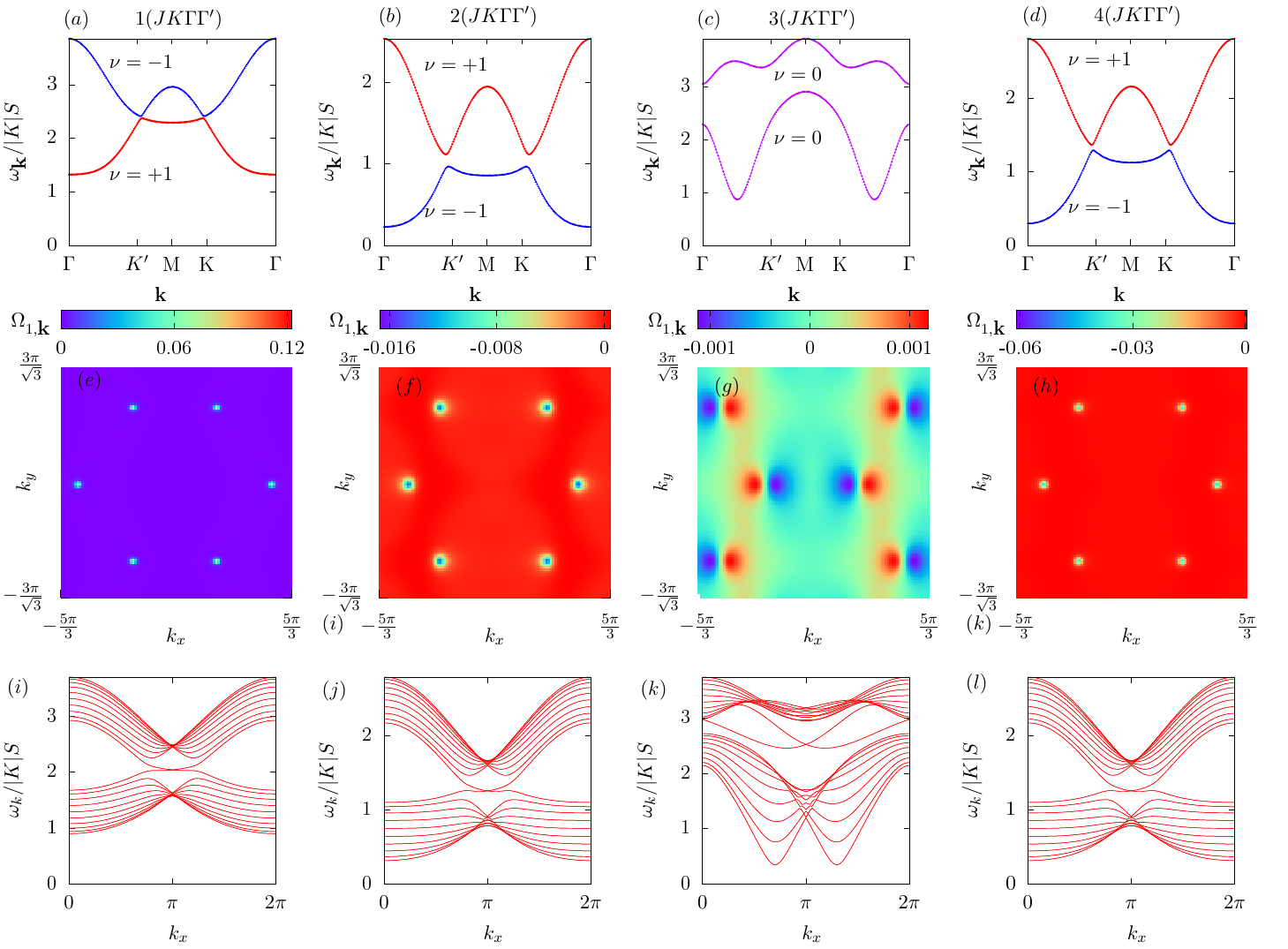}}
	\caption{ 
		Panels (a-d): Spin-wave bands, $\omega_{n\bf k}$, for $n=1,\;2$ (where $n$ denotes the band index), are depicted for the various models in Table~\ref{table:table1}. Panels (e-h): Berry curvature ($\Omega_{1{\bf k}}$) of the lower magnon band is presented.
		Panels (i-l) illustrate magnon bands along the zigzag edge of spin polarization in the $[11\bar{2}]$ direction. For all panels, an external magnetic field is applied parallel to the $[11\bar{2}]$ direction with the field strengths $h/S|K|$: 1.0 (first column),  0.1 (second column), 0.92 (third column), and 0.1 (fourth column).
	}
	\label{fig:fig4}
\end{figure}
As $\tau\rightarrow 0$ approaches zero, the non-negative weights exhibit a monotonic decrease with increasing energy~\cite{cao2015magnon}. Consequently, both the thermal Hall conductivity ($\kappa^{xy}_{\text{TH}}$) and the spin Nernst coefficient ($\kappa^{xy}_{\text{N}}$) vanish in the limit of zero temperature ($T\rightarrow 0$). This behavior arises because the temperature dependence of these transport coefficients is directly linked to the behavior of the weights.
At low temperatures, the contribution of low-energy magnons to the transverse thermal conductivities becomes more significant compared to high-energy magnons. This phenomenon arises because the lower magnon band achieves its maximum thermal occupancy at low temperatures.
Conversely, for systems characterized by a vanishing Chern number ($\nu=0$), the transport coefficients, $\kappa^{xy}_{\alpha}$ (where ${\alpha}$ represents both TH and N), exhibit markedly suppressed magnitudes compared to their counterparts in non-zero Chern number systems. However, these coefficients do not necessarily vanish entirely, and its sign is arbitrary.

Our study of the frustrated magnetic insulator $\alpha$-RuCl$_3$ commences with the premise that under a critical magnetic field, a high-field polarized state presents  its ground state configuration.
We focus on the fully ground state characterized by a finite minimum energy gap for a given set of exchange parameters $(J,K,\Gamma,\Gamma')$.
Fig.~\ref{fig:fig3} presents the calculated transverse thermal conductivities of the $(JK\Gamma\Gamma')$ model for the four distinct parameter sets detailed in Table~\ref{table:table1}.
These calculations were performed using linear spin wave theory, explore the influence of an applied magnetic field with varying orientations on the material's thermal transport properties.
Intriguingly, the sign structure demonstrates a pronounced sensitivity to two key factors: the orientation of the applied magnetic field and the exchange parameters of the nearest neighbor $(JK\Gamma\Gamma')$ to describe the $\alpha$-RuCl$_3$ material.

Our analysis of the $(JK\Gamma\Gamma')$ model, as presented in Fig.~\ref{fig:fig3}, reveals a striking dependence of the transverse thermal conductivities on the magnetic field orientation. When the external magnetic field (${\bf h}$) aligns along the [001] direction (Figs.~\ref{fig:fig3}a \& ~\ref{fig:fig3}e), all four proposed parameter sets consistently exhibit positive values.
This behavior aligns well with the dominance of thermally populated low-energy magnons within the lower magnon band, characterized by a Chern number of -1.
Conversely, a magnetic field oriented perpendicular to the honeycomb plane (${\bf h}||[111]$) induces a dramatic shift in behavior. All parameter sets enumerated in Table~\ref{table:table1} exhibit negative transverse thermal conductivities versus temperature (Figs.~\ref{fig:fig3}b \& ~\ref{fig:fig3}f). This observation is accompanied by a remarkable reversal of the Chern number for the lower magnon band, transitioning to a positive value for ${\bf h}||[111]$.
These findings may provide insights into the role of topology in thermal conductivity phenomena via the magnetic field dependence of low-energy magnons.
These contrasting behaviors highlight the intricate interplay between magnetic field orientation, topological character (Chern number), and thermal transport in this compound.
Furthermore, an accompanying in the thermal Hall signal was observed as the magnetic field strength increased. This reduction can be attributed to the widening of the magnon excitation gap, although this phenomenon is not explicitly illustrated in the current data visualization.

Now we proceed to evaluate the transverse thermal conductivities for the in-plane magnetic fields.
For ${\bf h}||[11\bar{2}]$, we calculate the temperature dependence of the in-plane thermal Hall conductivity, $\kappa^{xy}_{TH}$, for each model listed in Table~\ref{table:table1} using the SWT method. 
Our calculations for the models 1($JK\Gamma\Gamma'$), 2($JK\Gamma\Gamma'$), and 3($JK\Gamma\Gamma'$) reproduce the findings reported in Ref.~\cite{chern2024topological} (Fig.~\ref{fig:fig3}c), signifying good agreement.
Our theoretical analysis, focusing on the magnon-mediated contribution within the polarized phase of the model         4($JK\Gamma\Gamma'$), predicts a positive sign for the thermal Hall conductivity ($\kappa^{xy}_{\text{TH}}$) when the magnetic field is applied along the $[11{\bar 2}]$ direction.
This finding aligns remarkably well with the experimentally observed sign of $\kappa^{xy}_{\text{TH}}$ in $\alpha$-RuCl$_3$ (Ref.~\cite{czajka:2023}), signifying a strong concordance between theoretical predictions and experimental data.
However, it's important to note that this predicted positive sign for $\kappa^{xy}_{\text{TH}}$ along the ${\bf a}$ direction differs from the overall sign reported in Ref.~\cite{yokoi:2021half}.
As shown in Fig.~\ref{fig:fig3}c, within the ${\bf a}$-polarized state, the thermal Hall conductivity ($\kappa^{xy}_{\text{TH}}/T$) exhibits a magnitude of approximately $0.35 \times 10^{-3} {\rm mW}/{\rm K}^2{\rm m}$.
This value is the thermal Hall signals experimentally observed at low temperatures~\cite{Kasahara:2018,yokoi:2021half,czajka:2023,Imamura:2024}. For instance, the maximum value of $\kappa^{xy,2D}_{\text{TH}}/T$ for the model 4$(JK\Gamma\Gamma')$ is approximately $0.2\times (\pi k_B^2/6\hbar)$, constituting roughly $40\%$
of the half-quantized value.

Rotating the magnetic filed in the honeycomb plane from the ${\bf a}\parallel [11\bar{2}]$ direction toward the ${\bf b}\parallel [\bar{1}10]$ direction, our results show that the transverse thermal conductivities, $\kappa^{xy}_{\text{TH}}$ and $\kappa^{xy}_{\text{N}}$, begin to decrease and eventually to get a zero value in the field polarized state along the ${\bf b}$ direction (Figs.~\ref{fig:fig3}d \& ~\ref{fig:fig3}.
Under a magnetic field is applied along the crystallographic ${\bf b}$ direction, the Berry curvature exhibits antisymmetry with respect to $k_x=0$ (not shown).
This inherent antisymmetry property leads to a vanishing the integrands of expressions (\ref{eq:thermal}) and (\ref{eq:nernst}), consequently in zero transverse thermal conductivities.
Additionally, the magnon bands, characterized by a Chern number of zero, possess a topologically trivial character.
The zero thermal Hall conductivity of the polarized state arises from C$_2$ rotational symmetry around the ${\bf b}$ axis.
This symmetry enforces an antisymmetry in the Berry curvature, leading to vanishing integrands in the expression governing the transverse thermal conductivities (Eqs.~\ref{eq:thermal} \& \ref{eq:nernst}).
The sign structure of the thermal Hall conductivity ($\kappa_{\text{TH}}^{xy}$) arising from magnons in the field-polarized states reported in this study is consistent with that obtained in the Kitaev material $\alpha$-RuCl$_3$ when analyzed using a $K\Gamma\Gamma'$ model subject to in-plane magnetic fields~\cite{Chern:2021}. 
It is noteworthy that a pronounced anisotropy of magnonic properties with respect to the in-plane direction of an external magnetic field is indicative of a dominant Kitaev interaction~\cite{Zhang:2021}.

The thermal Hall conductivity is generally expected to vanish in a polarized state with a magnetic field applied along the ${\bf b}$ axis. However, this expectation can be circumvented by the presence of non-trivial magnetic orderings that break the $C_2$ symmetry. Such symmetry breaking allows for a finite thermal Hall effect even under the influence of a magnetic field aligned with the ${\bf b}$ direction~\cite{zhang2021topological}.
Recent thermal Hall conductivity measurements of the intermediate QSL phase have revealed a half-integer quantized plateau for magnetic fields applied along the ${\bf a}$ direction. Conversely, no such plateau was observed when the field was oriented along the ${\bf b}$ direction~\cite{Hwang:2022,Tanaka:2022,Imamura:2024}. As the magnetic field rotated within the honeycomb plane from ${\bf a}\parallel [11\bar{2}]$ to ${\bf b}\parallel [\bar{1}10]$, the thermal Hall conductivity ($\kappa_{xy}^{\rm TH}$) exhibited a sign change along the ${\bf b}$ direction. This change is indicative of a gap closure within the system, a phenomenon consistent with the $C_2$ rotational symmetry around the ${\bf b}$ direction.

\begin{figure}[t]
	\centerline{\includegraphics[width=.9\columnwidth]{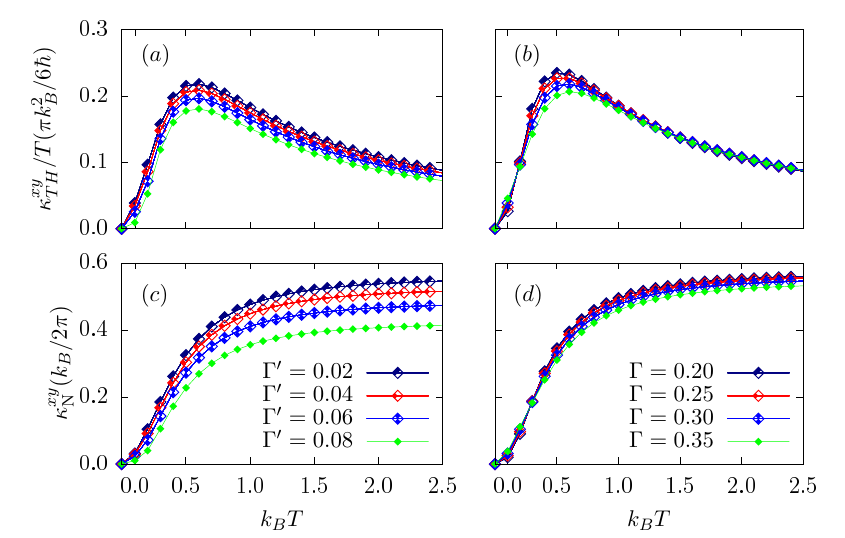}}
	\caption{
		Temperature dependence of $\kappa_{\text{TH}}^{xy}$ for the model 4$(JK\Gamma\Gamma')$ under a $[11\bar{2}]$ magnetic field, varying $\Gamma'$ (a), and $\Gamma$ (b).
		Temperature dependence of $\kappa_{\text{N}}^{xy}$ for the model 4$(JK\Gamma\Gamma')$ under a $[11\bar{2}]$ magnetic field, varying $\Gamma'$ (c), and $\Gamma$ (d).
	}
	\label{fig:fig5}
\end{figure}

\begin{figure}[t]
	\centerline{\includegraphics[width=1.\columnwidth]{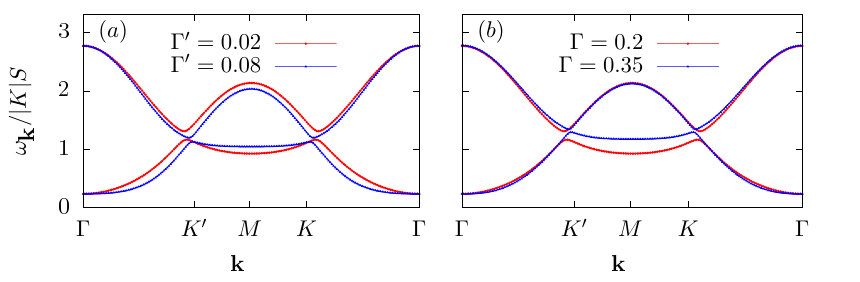}}
	\caption{ 
		Spin-wave dispersions ($\omega_{n{\bf k}}$) for the lower two bands ($n=1,~2$) of the model 4$(JK\Gamma\Gamma')$ under a $[11\bar{2}]$ magnetic field, varying $\Gamma'$ (a), and $\Gamma$ (b).		%
	}
	\label{fig:fig6}
\end{figure}

Here, we investigate the temperature-dependent behavior of the  transverse thermal conductivities in the $(JK\Gamma\Gamma')$ model with distinct parameter sets ($J,K,\Gamma,\Gamma'$) specified in Table~\ref{table:table1}.
This analysis focuses on elucidating the variations in the magnitude of the thermal Hall conductivity, $\kappa^{xy}_{\text{TH}}$, observed in Fig.~\ref{fig:fig3}c under the influence of an applied magnetic field along the $[11\bar{2}]$ direction. 
Our results in Fig.~\ref{fig:fig3}c reveal that the thermal Hall conductivity of the model 3($JK\Gamma\Gamma'$) is negligible, whereas the model 2($JK\Gamma\Gamma'$) exhibits a larger magnitude compared to the models 1($JK\Gamma\Gamma'$) and 4($JK\Gamma\Gamma'$).
Our findings for the model 3($JK\Gamma\Gamma'$) reveal an intriguing discrepancy. Although the Berry curvature of the model 3($JK\Gamma\Gamma'$) in momentum space exhibits a non-zero value (Fig.~\ref{fig:fig4}c), indicating a non-trivial character, the Chern number for the lower energy band remains zero ($\nu=0$).
The system exhibits a trivial topological phase characterized by a zero Chern number, resulting in the absence of topologically protected edge states (Fig.~\ref{fig:fig4}k).
This implies that the lower band is topologically trivial in the model 3($JK\Gamma\Gamma'$), and consequently, it does not contribute to the magnon thermal Hall conductivity ($\kappa^{xy}_{\text{TH}}$) upon the momentum integration in (\ref{eq:thermal}).
This observation underscores the critical role of a finite Chern number within the magnonic band structure for the emergence of a significant magnon thermal Hall effect.
Recent theoretical investigations have predicted the emergence of topological phases with Chern numbers exceeding $\nu=1$ within the ferromagnetic Kitaev-Heisenberg model, incorporating a Dzyaloshinskii-Moriya interaction. This model considers Heisenberg and Kitaev exchange terms on both nearest and next-nearest neighbor bonds, exploring a variety of configurations~\cite{Deb:2019,Deb:2020}.

We now delve into the origin of the disparity in the magnitude of $\kappa^{xy}_{\text{TH}}$ for the remaining three models.
At low temperatures, the lower magnon spectrum exhibits a finite excitation gap, resulting in a near zero response due to the lack of thermally excited magnons.
Upon increasing temperature within this low temperature regime, the observed variation in $\kappa^{xy}_{\text{TH}}$ can be primarily attributed to the contribution from the lower magnon band, particularly at small $k$-points.
This behavior is dictated by the Bose-Einstein distribution function.  A sufficiently small gap allows for thermal population of the lower band at low temperatures, thereby enabling a magnon contribution to $\kappa^{xy}_{\text{TH}}$.
We elucidate the role of the band gap in determining the magnitude of $\kappa^{xy}_{\text{TH}}$ by analyzing the models 2($JK\Gamma\Gamma'$) and 4($JK\Gamma\Gamma'$).
These models possess similar excitation gaps but exhibit distinct bulk gaps, as shown in Figs.~\ref{fig:fig4}b \& \ref{fig:fig4}d.
As evident in Fig.~\ref{fig:fig3}c, increasing temperature progressively suppresses the contribution from the lower magnon band to $\kappa^{xy}_{\text{TH}}$, ultimately leading to dominance by the lower band with a Chern number of opposite sign. 
Consequently, the band gap should not be too small. This ensures that the population of the upper band remains negligible across a wide temperature range.  Therefore, a significant contribution from the lower magnon band to the thermal conductivity, $\kappa^{xy}_{\text{TH}}$, is maintained.
Fig.~\ref{fig:fig3}g depicts the spin Nernst coefficient ($\kappa^{xy}_{\text{N}}$) for the models listed in Table~\ref{table:table1} under an external magnetic field aligned along the $[11\bar{2}]$ direction.
As discussed previously for the thermal Hall conductivity, these features stem from the non-trivial band structures of magnons.
Notably, the models 2$(JK\Gamma\Gamma')$ and 4$(JK\Gamma\Gamma')$ exhibit significant positive values of $\kappa^{xy}_{\text{N}}$ at low temperatures.
However, the coefficient exhibits a temperature dependence, vanishing at substantially higher temperatures, which are not presented here.

\begin{figure}
	\centerline{\includegraphics[width=0.85\columnwidth]{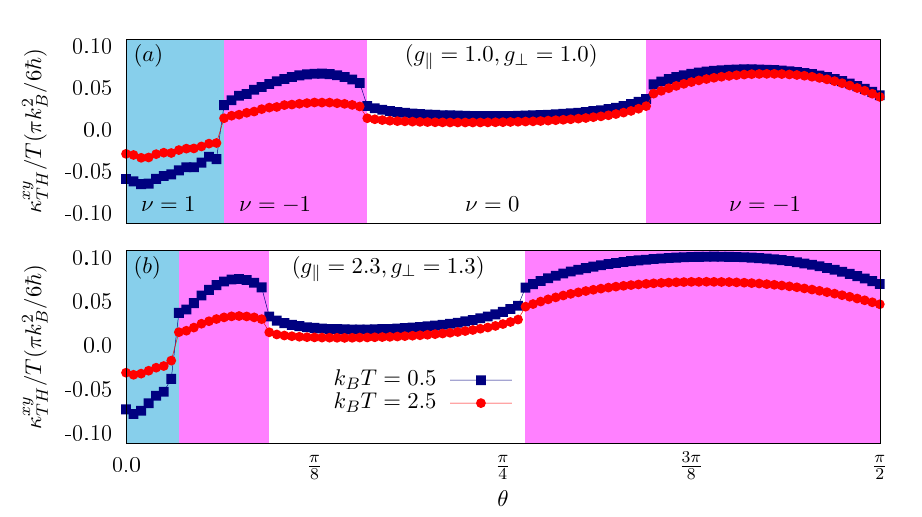}}
	\caption{ 		%
		Thermal Hall conductivity, $\kappa_{\text{TH}}^{xy}$, for the model 4$(JK\Gamma\Gamma')$ as a function of the tilting angle $\theta$ for two distinct cases: (a) an isotropic $g$-factor
		$(g_{\parallel}=1,g_{\perp}=1)$ with the field strength $h/S|K|=1.0$, and  (b) an anisotropic $g$-factor $(g_{\parallel}=2.3,g_{\perp}=1.3)$ with the field strength $h/S|K|=0.4$.
		Calculations are performed at various temperatures ($k_{\rm B}T=0.5,~2.5$).
	}
	\label{fig:fig7} 
\end{figure}

Recent theoretical studies suggest the possibility of experimentally tuning the exchange interactions in $\alpha$-RuCl$_3$ via octahedral distortion and layer stacking.
For example, compression could lead to enhanced overlap between neighboring atomic orbitals, resulting in anomalously large values of the off-diagonal exchange coupling term, $\Gamma$~\cite{luo2022interplay}.
These manipulations offer a promising approach for achieving exotic magnetic phases within the compound $\alpha$-RuCl$_3$~\cite{ahmadi2024ground}.
Motivated by the potential for engineering exotic magnetic phases in $\alpha$-RuCl$_3$ through the manipulation of exchange interactions, we investigate the role of the off-diagonal exchange interactions ($\Gamma,~\Gamma'$) for the magnitude and sign structure of the transverse thermal conductivities.
Henceforth, we concentrate on the model 4$(JK\Gamma\Gamma')$, as a comprehensive theoretical investigation of $\alpha$-RuCl$_3$ employing this specific parametrization is absent from the existing literature.
Figs.~\ref{fig:fig5} and ~\ref{fig:fig6} reveal a compelling relationship between the transverse thermal conductivities ($\kappa^{xy}_{\text{TH}}$ and $\kappa^{xy}_{\text{N}}$) and the band gap.
As the off-diagonal exchange interactions ($\Gamma$ and $\Gamma'$) increase, both $\kappa^{xy}_{\text{TH}}$ and $\kappa^{xy}_{\text{N}}$ values decrease in Fig.~\ref{fig:fig5}, mirroring the reduction in the band gap observed in Fig.~\ref{fig:fig6}.
This finding suggests that a non-negligible band gap is crucial for significant transverse thermal transport in this system.

Kitaev material candidate $\alpha$-RuCl$_3$ is known to possess a pronounced magnetic anisotropy, as documented by various experimental investigations~\cite{Johnson:2015,Sears:2015,sears:2020ferromagnetic,Li:2021b,Banerjee:2017,kubota2015successive,lampen2018anisotropic,weber2016magnetic}.
This anisotropic behavior is ascribed to a directionally dependent Land\'e $g$-factor~\cite{Johnson:2015,Li:2021b,kubota2015successive}. The Land\'e $g$-factor is a fundamental parameter quantifying a material's magnetic response.
In $\alpha$-RuCl$_3$, its anisotropy signifies a variation in the $g$-factor based on the applied magnetic field's orientation relative to the crystallographic axes.
This inherent anisotropic of the Land\'e $g$-factor is likely to be critical for understanding the behavior of the transverse thermal conductivities in $\alpha$-RuCl$_3$.

\begin{figure}[t]
	\centerline{\includegraphics[width=1.05\columnwidth]{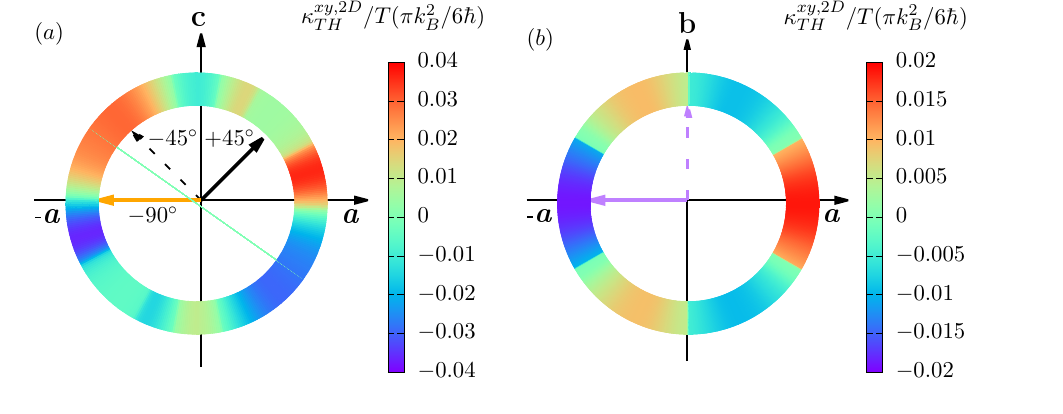}}
	\caption{ 
		(a) Variation of the thermal Hall conductivity ($\kappa_{\text{TH}}^{xy}$) for the 4$(JK\Gamma\Gamma')$ model under an external magnetic field. The field is applied within the $ac$ plane, with the solid and dashed black arrows indicating the field direction at $\theta = -45^\circ$ and $45^\circ$ relative to the $c$ axis,
		respectively. The orange arrow denotes the field direction
		along the $-a$ axis.
		(b) Variation of $\kappa_{\text{TH}}^{xy}$ in the presence of an external magnetic field within the $ab$ plane for the same model. The solid and
		dashed purple arrows represent the field direction along
		the $-a$ and $b$ axes, respectively.
		These calculations are performed at a temperature corresponding to $k_{\rm B}T=2.5$ and a field strength of $h/S|K|=5.5$, where $|K|$ represents the absolute value of the Kitaev exchange interaction. 
	}
	\label{fig:fig8}
\end{figure}

To elucidate the influence of magnetic anisotropy on the thermal Hall conductivity, we parameterize the applied magnetic field as ${\bf h}=hg_{\perp}\cos\theta{\hat{\bf c}}+hg_{\parallel}\sin\theta{\hat{\bf a}}$, where $\theta$ is the tilting angle defining the field's orientation relative to the $c$-axis (Fig.~\ref{fig:fig2}c).
We present the thermal Hall conductivity, $\kappa_{\text{TH}}^{xy}$, for the model 4$(JK\Gamma\Gamma')$ as function of $\theta$ for two distinct cases:
an isotropic $g$-factor $(g_{\parallel}=1,g_{\perp}=1)$ (Fig.~\ref{fig:fig7}a) and an anisotropic $g$-factor $(g_{\parallel}=2.3,g_{\perp}=1.3)$ (Fig.~\ref{fig:fig7}b).
These calculations are performed at various temperatures, ($k_{\rm B}T=0.5~{\rm K},~2.5~{\rm K}$).
Our calculations for the isotropic case (Fig.~\ref{fig:fig7}a) reveal a rich landscape of topological phase transitions as the tilting angle
$(\theta)$ varies from 0 to $\pi/2$.
Upon increasing $\theta$, the Chern number of the lower magnon band exhibits a positive value $(\nu=+1)$ at small tilt angels, transitioning to negative value $(\nu=-1)$ for larger angles.
An intermediate regime exists ($31\pi/200 < \theta < 63\pi/200$) where the Chern number of the lower magnon band vanishes, rendering the system topologically trivial.
Consequently, the thermal Hall conductivity exhibits a suppressed response in this regime. 
Finally, for tilt angles exceeding $63\pi/200$, the lower magnon band regains a non-trivial character, leading to a positive $\kappa^{xy}_{\text{TH}}$.
This sign reversal in these coefficients as a function of the tilting angle can be attributed to the interplay between the spatial distribution of the Berry curvature within FBZ and the temperature-dependent population of bands characterized by distinct Chern numbers. 
Therefore, the transverse thermal conductivities emerge as a powerful tool for elucidating the topological properties of magnons in magnetic insulators.
The anisotropic case (Fig.~\ref{fig:fig7}b) exhibits qualitatively similar phase transitions, albeit with distinct stability regions for each topological phase compared to the isotropic case.
Notably, Fig.~\ref{fig:fig7} demonstrates that the magnitudes of $\kappa^{xy}_{\text{TH}}$ for both isotropic and anisotropic cases are remarkably similar.

Building on the observation of a sign reversal in the thermal Hall conductivity ($\kappa^{xy}_{\text{TH}}$) within the spin liquid regime of $\alpha$-RuCl$_3$ as $\theta$ varied from $60^{\circ}$ to $-60^{\circ}$~\cite{yokoi:2021half}, we extend this investigation. 
Here, we explore the sign change of the magnon $\kappa^{xy}_{\text{TH}}$ within the model 4($JK\Gamma\Gamma'$) for an external field applied in both the $ac$-plane and $ab$-plane. This analysis is conducted assuming a partially-polarized ferromagnet (PPF) as the underlying magnetically ordered ground state.
Investigating the field angular variation the magnon $\kappa^{xy}_{\text{TH}}$ in the PPF state, a phase characterized by the absence of spontaneously symmetry breaking, akin to the QSL phase~\cite{yokoi:2021half}, is crucial for understanding the nature of the low-temperature excitations.
To guarantee the classical model's stability within the partially-polarized ferromagnet (PPF) phase, specific magnetic field strengths were chosen meticulously. 
Theoretical calculations predict a predominantly in-plane character of the magnetization upon entering the PPF phase. Crucially, complete polarization remains elusive even for finite fields applied in the $ac$-plane~\cite{zhang2021topological}.
This behavior arises from the interplay between the $\Gamma$ interaction and the external magnetic field~\cite{janssen2017magnetization}. As the field strength increases, the in-plane component of the magnetization exhibits a gradual decrease, while the out-of-plane component conversely experiences a gradual enhancement.

Fig.~\ref{fig:fig8}a depicts the field-angular dependence of $\kappa^{xy}_{\text{TH}}$ within the model 4($JK\Gamma\Gamma'$) for fields confined to the $ac$-plane.
Notably, we consider the scenario where only the in-plane component of a field tilted relative to the crystallographic $c$-axis is reversed, while the out-of-plane component remains unchanged.
In contrast to the QSL state previously observed in Ref.~\cite{yokoi:2021half}, our analytical results reveal that the magnon Hall conductivity $\kappa^{xy}_{\text{TH}}$ does not exhibit a sign change and equivalent magnitude at $\theta=\pm 60^{\circ}$ within the PPF phase.
This asymmetry exclusively emerges when the Zeeman energy is predominantly dictated by the in-plane field component. This behavior is a direct consequence of the previously established observation: upon entering the PPF phase, the magnetization adopts a predominantly in-plane character~\cite{zhang2021topological}.
The magnon $\kappa^{xy}_{\text{TH}}$ of the PPF state exhibits a sign reversal upon field reversal, irrespective of the initial field orientation. However, the magnitude of $\kappa^{xy}_{\text{TH}}$ remains invariant under such reversal.

In the case of purely in-plane magnetization, a hallmark of the PPF phase, Fig.~\ref{fig:fig8}b reveals a sign reversal in $\kappa^{xy}_{\text{TH}}$. Intriguingly, the magnitudes of $\kappa^{xy}_{\text{TH}}$ for both in-plane field orientations (along the $a$ and $-a$ directions) are mirrored, exhibiting opposite signs. This behavior aligns perfectly with the theoretical predictions outlined in Ref.~\cite{Chern:2021}.
Consequently, our findings strongly support the notion that the sign change of $\kappa^{xy}_{\text{TH}}$ remains prevalent as long as the in-plane field component retains dominance over the Zeeman energy.
We demonstrated a sign reversal of the thermal Hall conductivity ($\kappa^{xy}_{\text{TH}}$) for low-energy magnon excitations in the purely in-plane magnetized PPF phase. This aligns with experimental findings for Majorana excitations in non-Abelian Kitaev spin liquids~\cite{yokoi:2021half,Imamura:2024}.
However, this sign structure alone is insufficient to definitively confirm the presence of Majorana fermions in the Kitaev spin liquid. A more conclusive test lies in the quantization of $\kappa^{xy}_{\text{TH}}$ at very low temperatures, as magnon contributions vanish at this limit. Recent experimental studies have demonstrated that the angular dependence of low-temperature specific heat near the b direction can effectively distinguish between itinerant Majorana fermions and other excitations responsible for the planar thermal Hall effect in $\alpha$-RuCl$_3$~\cite{Imamura:2024}.

\section*{Conclusion}

In summary, this study provides a theoretical investigation of the topological properties of low-energy magnetic excitations in $\alpha$-RuCl$_3$, a promising Kitaev material. 
The non-trivial topology in this system results in the emergence of transverse thermal transport phenomena, specifically the thermal Hall conductivity and spin Nernst effect, when subjected to an external magnetic field.
In the high-field polarized regime, we assess the effectiveness of various proposed models for $\alpha$-RuCl$_3$ (see Table~\ref{table:table1}) in explaining the experimentally observed thermal Hall conductivity at low-temperatures, which is attributed to topological magnons.
In contrast to the sign of $\kappa_{\text{TH}}^{xy}$ in the non-Abelian Kitaev spin liquid, which is exclusively determined by the field direction~\cite{Kitaev:2006}, our findings reveal that the sign of $\kappa_{\text{TH}}^{xy}$ in the PPF phase is influenced by a combination of factors: the orientation of the applied magnetic field and the nearest-neighbor exchange parameters $JK\Gamma\Gamma'$ used to model $\alpha$-RuCl$_3$.
Meanwhile, the obtained results suggest that the presence of three crucial conditions is necessary for the achievement of notable magnon thermal conductivity: (1) the existence of topological magnonic bands, (2) a limited excitation gap that facilitates thermal population of the lower band, and (3) a substantial band gap that minimizes occupation of the upper band across different temperature ranges.
In contrast to the spin liquid regime, where the thermal Hall conductivity exhibits a sign reversal as $\theta$ varies from $60^{\circ}$ to $-60^{\circ}$, our findings indicate that the sign change of the thermal Hall conductivity within the PPF state prevalent as long as the in-plane field component maintains dominance over the Zeeman energy.
We have observed a sign reversal in the thermal Hall conductivity for magnons within the PPF phase, which is the purely magnetized in-plane.
This phenomenon bears resemblance to the behavior of Majorana excitations found in Kitaev spin liquids. Nonetheless, it is important to note that this observation alone does not suffice to establish the presence of Majorana fermions.
To obtain more definitive evidence, further investigations are needed, including the examination of the half-integer quantization of $\kappa_{\text{TH}}^{xy}$ at low temperatures and the analysis of the angular dependence of specific heat~\cite{Imamura:2024}.

\section*{Data Availability}
The data that support the findings of this study are available within the article.

\section*{Appendix: Matrices Elements of the SWT Hamiltonian}
\label{appA}

The momentum-dependent functions ${\bf {A(k)}}$ and ${\bf {B(k)}}$ in (\ref{eq:matrixDk}) are given by:

\begin{equation}
	{\bf {A(k)}} =
	\left(
	\begin{matrix}
		\lambda_{0} & \lambda_{1,\kk} \\
		\lambda_{1,\kk}^{*} & \lambda_{0}
	\end{matrix}
	\right) \,; ~~
	{\bf {B(k)}} =
	\left(
	\begin{matrix}
		0 & \lambda_{2,\kk} \\
		\lambda_{2,-\kk} & 0
	\end{matrix}
	\right) \,,
\end{equation}
with

\begin{align}
	\begin{aligned}
		{\lambda_0} & =
		\begin{cases}
			-3J-K+\frac{h}{S}, & \quad \text{for } {\bf h} \parallel [001], \\
			-3J-K-2\Gamma-4\Gamma'+\frac{h}{S}, &  \quad \text{for } {\bf h} \parallel [111], \\
			-3J-K+\Gamma+2\Gamma'+\frac{h}{S}, &
			\quad \text{for } {\bf h} \parallel [11{\bar 2}],
			\\
			-3J-K+\Gamma+2\Gamma'+\frac{h}{S}, & \quad \text{for } {\bf h} \parallel [\bar{1}10],  
		\end{cases}
	\end{aligned}
\end{align} 
and
\begin{align}
	\begin{aligned}
		{\lambda}^{}_{1,\bf k} & =
		\begin{cases}
			J+(J+\frac{K}{2})\Big(e^{-i\kk_{1}}+e^{-i\kk_{2}}\Big), & \quad \text{for } {\bf h} \parallel [001], \\
			(J+\frac{K}{3}-\frac{\Gamma}{3}-\frac{2\Gamma'}{3})\Big(1+e^{-i\kk_{1}}+e^{-i\kk_{2}}\Big), & \quad \text{for } {\bf h} \parallel [111], \\
			(J+\frac{K}{6}-\frac{\Gamma}{6}+\frac{2\Gamma'}{3})+	(J+\frac{5K}{12}+\frac{2\Gamma}{3}+\frac{\Gamma'}{6})\Big(e^{-i\kk_{1}}+e^{-i\kk_{2}}\Big), & \quad \text{for } {\bf h} \parallel [11{\bar 2}],
			\\
			K(\frac{e^{-i\kk_{1}}+e^{-i\kk_{2}}}{4}+\frac{1}{2})+\frac{\Gamma'}{2}\Big(e^{-i\kk_{1}}+e^{-i\kk_{2}}\Big), & \quad \text{for } {\bf h} \parallel [\bar{1}10],  
		\end{cases}
	\end{aligned}
\end{align} 
where $\kk_{1,2}=\kk\cdot{\bf a}_{1,2}$ with  ${\bf a}_{1,2}=a(\pm 1/2, \sqrt{3}/2)$  ($a$ denotes the
lattice distance between two nearest neighbor sites on the honeycomb lattice, and we shall set $a = 1$). Furthermore, 

\begin{align}
	\begin{aligned}
		{\lambda}^{}_{2,\bf k} & =
		\begin{cases}
			i\Gamma+\frac{K}{2}\Big(e^{-i\kk_{1}}-e^{-i\kk_{2}}\Big), & \quad \text{for } {\bf h} \parallel [001], \\
			(\frac{K}{3}+\frac{2\Gamma}{3}-\frac{2\Gamma'}{3})\Big(1+e^{-i(\kk_{1}+\frac{2\pi}{3})}+e^{-i(\kk_{2}+\frac{2\pi}{3})}\Big), & \quad \text{for } {\bf h} \parallel [111], \\
			(\frac{K}{6}+\frac{5\Gamma}{6}+\frac{2\Gamma'}{3})+	\frac{i}{\sqrt{6}}(K-\Gamma+\Gamma')
			+(-\frac{K}{12}+\frac{\Gamma}{3}+\frac{7\Gamma'}{6})\Big(e^{-i\kk_{1}}+e^{-i\kk_{2}}\Big), & \quad \text{for } {\bf h} \parallel [11{\bar 2}],
			\\
			K\Big(-\frac{e^{-i\kk_{1}}+e^{-i\kk_{2}}}{4}+\frac{1}{2}\Big)-\frac{\Gamma'}{2}\Bigg((1+i\sqrt{2})\Big(e^{-i\kk_{1}}+e^{-i\kk_{2}}\Big)+2i\sqrt{2}\Bigg), & \quad \text{for } {\bf h} \parallel [\bar{1}10],  
		\end{cases}
	\end{aligned}
\end{align} 

\bibliography{References}

\end{document}